\documentclass[twocolumn,final]{IEEEtran}
\usepackage{graphicx}
\usepackage{bm}
\usepackage{url,mystyle}
\usepackage{amsfonts,amssymb,amsmath}
\renewcommand\url{\begingroup\urlstyle{rm}\Url}
\centerfigcaptionstrue
\allowdisplaybreaks

\title{Relay-Assisted Interference Network: Degrees of Freedom}
\author{Ramy Abdallah Tannious and Aria
  Nosratinia, {\em Fellow, IEEE}
\thanks{Manuscript received October 2008; revised August 2011,
  accepted December 2011. This work was supported in part by the
  National Science Foundation under grant CNS-0435429. The material in
  this paper was presented in part at the 2008 IEEE International
  Symposium on Information Theory (ISIT), Toronto,
  Canada.}\thanks{Ramy Abdallah Tannious was with The University of
  Texas at Dallas, he is now with Aviat Networks, Santa Clara, CA,
  95054, USA, e-mail: ramy@alumnimail.utdallas.edu. Aria Nosratinia is
  with the Department of Electrical Engineering, The University of
  Texas at Dallas, Richardson, TX 75083, USA, e-mail:
  aria@utdallas.edu.}}

\begin{document}

\maketitle

\begin{abstract}
This paper investigates the degrees of freedom of the interference
channel in the presence of a dedicated MIMO relay. The relay is used
to manage the interference at the receivers. It is assumed that all
nodes including the relay have channel state information only for
their own links and that the relay has $M \geq K$ antennas in a
$K$-user network. We pose the question: What is the benefit of
exploiting the direct links from the source to destinations compared
to a simpler two-hop strategy.  To answer this question, we first
establish the degrees of freedom of the interference channel with a
MIMO relay, showing that a $K$-pair network with a MIMO relay has
$\frac{K}{2}$ degrees of freedom.  Thus, appropriate signaling in a
two-hop scenario captures the degrees of freedom without the need for
the direct links. We then consider more sophisticated encoding
strategies in search of other ways to exploit the direct links.  Using
a number of hybrid encoding strategies, we obtain non-asymptotic
achievable sum-rates. We investigate the case where the relay (unlike
other nodes) has access to abundant power, showing that when sources
have power $P$ and the relay is allowed power proportional to
$\mathcal{O}(P^2)$, the full degrees of freedom $K$ are available to
the network.
\end{abstract}


\section{Introduction}

In addition to historical significance in network information theory, a
better understanding of the interference channel~\cite{Carleial78} is
becoming increasingly practically important, since many current wireless
communication systems are interference-limited. Examples include ad-hoc
networks with peer-to-peer communications that lack infrastructure and
hence transmission coordination, interference between adjacent networks
in wireless LAN systems, as well as cognitive networks, where primary
and secondary users transmit in the same band.

The capacity of the interference channel in the most general case
remains unknown, thus a number of partial approaches for investigating
the interference channel have been pursued. One of the tools for
understanding the behavior of multi-terminal networks is the \emph
{degrees of freedom} (DOF), also known as the \emph {multiplexing gain}
or the \emph{ pre-log factor}, which characterizes the scaling behavior
of a network throughput at high signal-to-noise ratios (SNR). We
formally define the degrees of freedom as follows~\cite{Azarian05}:
\begin{equation}
DOF= \lim_{P\to\infty} \frac{C_s}{\log(\frac{P}{\sigma^2})}
\label{eq:DOF}
\end{equation}
where $P$ is the power constraint at each source node, $\sigma^2$ is
the noise variance at a destination and $C_s$ is the network sum-rate
capacity. For example, the maximum degrees of freedom of a two-user
(single-antenna) Gaussian interference channel is equal to
one~\cite{Host-Madsen06}.  

This work investigates the effect of having a dedicated MIMO relay
shared by several source-destination pairs on the degrees of freedom
of such network.  The main issue is whether with simple single-user
decoding at the destinations, exploiting direct links is of a benefit.

Recent advances in network information theory have led to the
characterization of the degrees of freedom of several networks. It is
well known that the MIMO MAC and MIMO BC have full degrees of
freedom~\cite{Tse04,Vishwanath03}. Thus, the degrees of freedom in the
MIMO MAC and BC channels do not increase with transmit and receive
cooperation, respectively.

Recently, the phenomenon of interference alignment has led to new
results that characterize the degrees of freedom in various
interference networks.  The idea of interference alignment is for the
transmissions to coordinate in such a manner such that at the
receivers the interference signals overlap in certain dimensions and
therefore other dimensions are left interference-free. Via
interference alignment, in a $K$-user time-varying interference
network $\frac{K}{2}$ degrees of freedom are achieved almost
surely~\cite{Jafar08b}. 

The first attempt to study the effect of relaying on the degrees of
freedom of the interference network was performed
in~\cite{Host-Madsen06} and~\cite{Host-Madsen05}. A rather negative
result was obtained, showing that cooperation over fading links
between the sources, between the destinations, or both, cannot improve
the degrees of freedom of an interference network. On the other hand,
if perfect cooperation between sources (destinations) is assumed, the
network can mimic a MIMO system with antennas co-located at the
transmitting (receiving) side as mentioned previously.
In~\cite{Ng07}, the links between sources or between destination are
considered having phase fading and it is shown that cooperation can
help in increasing the throughput of a two-user interference channel
close to rates achieved by a $2\times 2$ MIMO system.  Considering
distributed dedicated relays, Morgenshtern and
B\"olcskei~\cite{Boelcskei07} showed that the interference network can
decouple. This is based on devising an amplify-and-forward two-hop
strategy that utilize full (but local) CSI at the relays and subject
to having the number of relays $m$ greater that $K^3$, where $K$ is
the number of source-destination pairs.  A similar decoupling can be
achieved by fewer relays, specifically $m \geq K^2$ with the cost of
having global CSI at the relays~\cite{Hassibi06}. The DMT performance
of this scheme was further analyzed in~\cite{Raophd}.  Finally, a
two-hop network with MIMO relaying decouples into a MIMO-MAC followed
by a MIMO-BC each achieving full degrees of freedom~\cite{Jafar08},
therefore the DoF $\frac{K}{2}$ is achievable with a two-hop
transmission.

The addition of a MIMO relay to an interference channel (including
direct links) gives rise to a network model that we denote {\em the
  interference MIMO relay channel (IMRC)}. In this paper we first
establish the degrees of freedom of the Gaussian IMRC with the source
and destination nodes having one antenna each. Achievability is
demonstrated with a two-hop scheme, without exploiting the direct
links from the sources to the destinations. The upper bound on the
degrees of freedom is obtained by specializing the recently developed
upper bounds in~\cite{Jafar08c}. We establish that the interference
MIMO relay channel has $\frac{K}{2}$ degrees of freedom.

We then take the investigation one step further to consider degrees of
freedom beyond $\frac{K}{2}$.  We devise new combinations of coding
strategies that are inspired by the coding schemes used in relay
channels, as well as MIMO MAC and MIMO broadcast channels. These
coding strategies attempt to exploit the direct links but at the same
time manage the interference at the receivers using the MIMO relay. It
is assumed that all nodes, including the relay, have only their own
channel state information.  We further consider the effect of the
availability of abundant power at the relay.  This is motivated by
real-world scenarios where a single relay tower, with easy access to
power, is assisting many mobiles. We wish to understand whether the
devised coding scheme and the additional power at the relay can
improve the DOF of the channel. Also, we investigate the minimum
amount of power needed to impart maximum degrees of freedom to the
network. We find that if the relay has $M \geq K$ antennas and power
proportional to $\mathcal{O}(P^2)$, it can impart the maximum $K$
degrees of freedom to a $K$-user network whose users have power
$\mathcal{O}(P)$, regardless of the number of users (c.f. of our
definition of DOF in~(\ref{eq:DOF})).

The remainder of this paper is organized as follows.
Section~\ref{sec:system} explains the notations used in the paper and
provides the system model.  Section~\ref{sec:DOF} establishes the DOF
of the $K$-user IMRC.  Section~\ref{sec:coding} states the main result
of the paper and presents the detailed coding strategies that exploits
the direct links and abundant power at the relay. We corroborate our
analytical findings by numerical results in
Section~\ref{sec:simulations}. Finally, Section~\ref{sec:conc}
concludes the paper.

\section{System Model}
\label{sec:system}

Throughout the paper, lower-case and upper-case boldface letters
denote vectors and matrices, respectively. The determinant of matrix
${\bf X}$, it's transpose and Hermitian are denoted $\det({\bf X})$,
${\bf X}^\dagger$ and ${\bf X}^*$, respectively. The norm of a vector
${\bf x}$ is denoted by $\|{\bf x}\|$. $\log(\cdot)$ stands for the
base-2 logarithm. All rates are expressed in bits/channel use.

\begin{figure}
\centering
\includegraphics{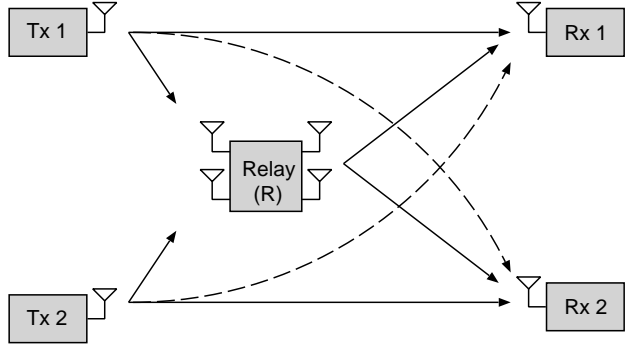}
\caption{The Two-User Interference MIMO Relay Channel}
\label{fig:IMRC}
\end{figure}

The \emph{interference MIMO relay channel} (IMRC) is depicted in
Fig.~\ref{fig:IMRC}. Nodes $1$ and $2$ attempt to communicate
independent messages $W_1$ and $W_2$ to their respective receivers,
possibly with help from the relay (node $R$).  The relay is assumed to
be equipped with $M$ antennas, where $M\geq 2$ while all other nodes
have one antenna each. All links are subject to flat fading which
remains constant during the transmission period. The channels from the
sources to their corresponding destinations, from the sources to the
relay and from the relay to destinations are denoted by the letters,
$f$, $g$ and $h$, respectively. A subscript $ab$ is used to index the
transmitting and receiving nodes, $a$ and $b$, respectively.

The input-output relation of a Gaussian IMRC is given by:
\begin{align}
y_1&=f_{11}x_1+ {\bf h}_{R1}^{\dagger} {\bf x}_R+ f_{21}x_2+z_1\label{eq:rec3}\\
y_2&=f_{12}x_1+ {\bf h}_{R2}^{\dagger}{\bf x}_R+ f_{22}x_2+z_2\label{eq:rec4}\\
y_R&={\bf g}_{1R}x_1+ {\bf g}_{2R}x_2+z_R
\end{align}
where $y_1$, $y_2$ and $y_R$ are the channel outputs at receivers $1$,
$2$ and the relay, $x_1$, $x_2$ and ${\bf x}_R$ are the transmitted
signals. The variables $z_1$, $z_2$ and $z_R$ denote zero-mean,
unit-variance additive white Gaussian noises at the receivers. We
assume individual block power constraints on the transmitting
nodes. Nodes $1$ and $2$ have equal transmit power constraint of $P$,
i.e.
\begin{equation}
\sum_{i=1}^{n}\|x_k(i)\|^2 \leq nP, \qquad\qquad k=1,2
\end{equation}
where $i$ is the symbol index within a block of $n$ symbols.

We assume that the relay node has block power constraint $P_R$, which
may be different from $P$ and will be specified in each instance in
the sequel. The relay uses a decode-and-forward scheme~\cite{Cover79}
that includes linear pre-coding, in a manner to be explained
shortly. The channel state information (CSI) knowledge assumptions are
as follows. Transmitters $1$ and $2$ each have perfect knowledge about
their own transmit-side CSI while receivers $1$ and $2$ have perfect
knowledge of their receive-side CSI. The relay is assumed to have
knowledge of its incoming and outgoing links. The relay is assumed to
operate in full-duplex mode, i.e., it can receive and transmit at the
same time. Throughout this paper, we assume the input alphabets to be
Gaussian. The average probability of error is defined as follows:
\begin{equation}
P_e^{(n)}= Pr\big[\{\hat{W}_1\neq W_{1}\} \cup \{\hat{W}_2\neq
W_2\}\big]
\end{equation}
where, $\hat{W}$ denotes an estimate of $W$.  The rate of transmission
from node $k$ is $R_k=\frac{\log Q_k}{n}$, where $Q_k$ is the size of
the message transmitted by node $k$. A rate pair $(R_1,R_2)$ is
said to be achievable for the interference MIMO relay channel if there
exist a sequence of codes
$\big(\big(2^{nR_{1}},2^{nR_{2}}\big),n\big)$ with average probability
of error $P_{e}^{(n)}\rightarrow 0$ as $n \rightarrow \infty$. A
$K$-user interference network with a single MIMO relay can be defined
as a straightforward extension to the above model.

\section{The Degrees of Freedom of IMRC}
\label{sec:DOF}

The first main result of this paper is as follows:

\begin{theorem}
\label{theorem:DOF}
 The degrees of freedom of the interference MIMO relay network is
 $\frac{K}{2}$.
\end{theorem}

\begin{Proof}
Achievability is established with a simple two-hop scheme. The first
phase where the sources transmit the signals and the MIMO relay
decodes is a MIMO MAC channel. It is well known that this channel
achieves the full $K$ degrees-of-freedom. The second phase where the
relay transmit to the receivers is a MIMO BC and again is know to
achieve $K$ DOF. The transmission in two hops entails a penalty of one
half in the DOF. Thus, $\frac{K}{2}$ DOF are achieved for the IMRC.

Now, the converse. A recent work~\cite{Jafar08c} produced an elegant
approach to find upper bounds on fully connected interference and X
networks with relays and feedback. The upper bound on a $S\times R
\times D$ fully connected network can be specialized to the network we
study in this paper where $S$, $R$ and $D$ refer to the number of
sources, relays and distentions in the network. A fully connected
network means that there is a message from every source to every
destination. For completeness, we will first state the main result on
the upper bound on the degrees of freedom of the $S\times R \times D$
network.

\begin{theorem}~\cite{Jafar08c}
If $\mathcal{D}$ represents the degrees of freedom region of the
$S\times R \times D$ node $X$ network, then the total degrees of
freedoms can be upper bounded as follows:
\begin{equation*}
\max_{[(d_{i,j})]\in \mathcal{D}} \sum_{j=1}^{S}\sum_{i=S+R+1}^{S+R+D} d_{i,j} \leq \frac{SD}{S+D-1}
\end{equation*}
\end{theorem}
Note that~\cite{Jafar08c} derives upper bound not only on the degrees
of freedom of the $S\times R \times D$ but on the whole degrees of
freedom \emph{region}. The interested reader is refered
to~\cite{Jafar08c} for further details.

Now for the $K$-user interference network, using the following
corollary from~\cite{Jafar08c} the exact degrees of freedom is
obtained.
\begin{corollary}
Consider a fully connected $K$ user interference network with $R$
  relays, where all the channel coefficients are
  time-varying/frequency-selective with values drawn randomly from a
  continuous distribution with support bounded below by a non-zero
  constant. Let all nodes be full-duplex allowing noisy
  transmitter/receiver cooperation. Also, let the source and relay
  nodes receive perfect feedback from all nodes. Then the interference
  network has $\frac{K}{2}$ degrees of freedom.
\end{corollary}
The bounds in the previous theorem and corollary are applicable to the
MIMO relay in the IMRC, because the proof of the converse assumes full
cooperation between the distributed $R$ relay nodes (see
observation~$3$ in~\cite{Jafar08c}). Also, feedback and time/frequency
selectivity of the channel do not reduce the degrees of freedom of the
channel. Therefore, due to the matching achievability and converse
results, the DOF of IMRC is established to be $\frac{K}{2}$.
\end{Proof}

\section{Degrees of Freedom Beyond $K/2$}
\label{sec:coding}

In the previous section, the DOF of the interference channel with a
MIMO relay was shown to be $\frac{K}{2}$, which is also achievable via
a two-hop strategy. Therefore, the direct links do not contribute to
the DOF of the channel. However, we will show that the DOF can be
larger than $\frac{K}{2}$ by exploiting the direct links in addition
to a more powerful MIMO relay.

We start by developing a coding strategy that uses the direct links
and investigate, through the derived sum rate of the channel, the
reason for the inefficiency of the direct links in improving the
DOF. Then, the effect of abundant power at the relay on the DOF is
studied.

\subsection{Coding Strategies and Achievable Rates}\label{subsec:strategy}

The idea of the upcoming coding strategies is to use the relay in a way that
minimizes the interference at the receivers. This task is
highly nontrivial because the causality of the relay prohibits
straight-forward interference cancelation. Therefore, sophisticated
coding and power control strategies are needed to possibly manage the
interference at the receivers.

Consider a transmission period of $B$ blocks, each of $n$ symbols. It
is assumed that $n$ is sufficiently large to allow reliable
decoding. Without loss of generality, at first a two-user network is
considered. Nodes $1$ and $2$ send sequences of $B-1$ messages
$(W_1(b)$ and $W_2(b))$, respectively, over the channel in $nB$
transmissions, where $b$ denotes the block index, $b = 1,
2,\ldots,B-1$. The rate pair $(R_1\frac{B-1}{B},R_2\frac{B-1}{B})$
approaches $(R_1,R_2)$ as $B\rightarrow\infty$.
\subsection{Encoding at the Sources}
The source uses the super-position block Markov encoding technique
devised in~\cite{Cover79}. In particular at any block $b$,
\begin{align}
X_1^{(b)}&= U_1+ U_1'\\
X_2^{(b)}&= U_2+ U_2'
\end{align}
where $U_1$ and $U_1'$ are i.i.d Gaussian codebooks encoding the
messages of the current and the previous blocks with powers $\chi(P)$
and $\psi(P)$, respectively, according to the power constraint
\begin{equation}\label{powerconst}
\chi(P)+\psi(P)=P
\end{equation}

Similar definitions hold for the signal components transmitted by node $2$, $U_2$ and $U_2'$.

\subsection{Decoding and Re-encoding at the Relay}

A space division multiple-access (SDMA) approach is used to
communicate between nodes $1$, $2$ and the MIMO relay. Therefore, both
sources transmit simultaneously and the MIMO relay attempts decoding
both signals.  At the end block $b$, given that the relay decoded both
messages $W_1(b-1)$ and $W_2(b-2)$ correctly, it can decode the
messages $W_1(b)$ and $W_2(b)$ of both users while achieving a
$DOF=2$. This can be achieved by a zero-forcing strategy, as long as
the relay has no fewer antennas as the number of transmit nodes, and
is made possible by the independence of the users' channels to the
relay that is a result of spatial separation. The sum-rate constraint
for correct decoding at the relay is given by~\cite[Section
  10.1]{Tsebook}:
\begin{equation}
\label{eq:CsMAC}
R_1+R_2\leq \log\det\big({\bf I}_2 + {\bf GK}_x{\bf G}^*\big)
\end{equation}
where ${\bf G}=[{\bf g}_{1R}\,\, {\bf g}_{2R}]$, ${\bf
  K}_x=diag\big(\chi(P),\chi(P)\big)$, and ${\bf I}_2$ is the
$2\times2$ identity matrix.

Now, to the relay encoding strategies. Ideally, it would be desirable
for the relay to cancel the entire interference at each receiver.
However, due to causality, the relay can only cancel the interference
arising from signals that it has already decoded. Thus, even if
everything is accomplished perfectly, not all of the interferences
will be canceled. The question is, if interferences cannot be fully
removed, then how must the remaining interference be managed so that a
good result may be obtained in terms of the degrees of freedom. This
issue will be addressed in the sequel via power allocation policies at
the sources and at the relay.

The channel from the relay to both destinations is similar to a
Gaussian MIMO broadcast channel whose capacity region has been
recently determined~\cite{Shamai07}. To help in canceling the
interference, the relay uses a modified zero-forcing beamforming
(ZF-BF) strategy~\cite{Goldsmith06}. ZF-BF achieves the maximum
degrees of freedom of the sum-rate capacity of a Gaussian MIMO BC,
although it is in general suboptimal compared to the
capacity-achieving dirty-paper coding (DPC) strategy. The relay
constructs and transmits the following signal:
\begin{equation}
{\bf x}_R^{(b)}=u_1'{\bf t_1}+u_2'{\bf t_2}
\end{equation}
where ${\bf t}_1$ and ${\bf t}_2$ are $2 \times 1$ complex beamforming
vectors. For simplicity, we assume the relay divides its power $P_R$
equally between the two signals components, i.e. $||{\bf t}_1||^2 =
||{\bf t}_2||^2 = \frac{P_R}{2\psi(P)}$.  Proper selection of
beamforming vectors (magnitudes and phases) allows partial suppression
of interference at the receivers as will be described later.

\subsection{Decoding at the Destinations}

Given the structure of the signal formed by the relay, we re-write
(\ref{eq:rec3}) and (\ref{eq:rec4}) as follows:
\begin{align}
y_1^{(b)}=&f_{11}u_1+ (f_{11}+{\bf h}_{R1}^{\dagger}{\bf t}_1)u_1'+
(f_{21}+{\bf h}_{R1}^{\dagger}{\bf t_2})u_2'\nonumber\\
&\qquad +f_{21}u_2+z_1\label{eq:modrec3}\\
y_2^{(b)}=&f_{12}u_1+ (f_{12}+{\bf h}_{R2}^{\dagger}{\bf t}_1)u_1'+
(f_{22}+{\bf h}_{R2}^{\dagger}{\bf t_2})u_2'\nonumber\\
&\qquad +f_{22}u_2+z_2\label{eq:modrec4}
\end{align}
Therefore, the beamforming vectors at the relay ${\bf t}_1$ and ${\bf
  t}_2$ are selected such that ${\bf h}_{R2}^{\dagger}{\bf
  t}_1=-f_{12}$ and ${\bf h}_{R1}^{\dagger}{\bf t}_2=-f_{21}$. The
derivation of ${\bf t}_1$ and ${\bf t}_2$ is discussed in the
Appendix.  This will cancel part of the interference seen by each
receiver, thus the received signals are modified to:
\begin{align}
y_1^{(b)}=&f_{11}u_1+ (f_{11}+{\bf h}_{R1}^{\dagger}{\bf t}_1)u_1'+f_{21}u_2+z_1\label{eq:finalrec3}\\
y_2^{(b)}=&f_{12}u_1+ (f_{22}+{\bf h}_{R2}^{\dagger}{\bf t_2})u_2'+f_{22}u_2+z_2\label{eq:finalrec4}
\end{align}
Receivers $1$ and $2$ can use Willems's backward decoding to decode
their intended signals~\cite{Willems82}. Backward decoding imposes
decoding delays, however, it simplifies the analysis compared to list
decoding or window decoding~\cite{Kramer05}. Backward decoding starts
from block $B$. The receivers have interference-free channels to
decode $u_1^{(B-1)}$ and $u_2^{(B-1)}$. In block $B-1$, they
pre-subtract the components of $u_1^{(B-1)}$ and $u_2^{(B-1)}$ before
attempting to decode $u_1^{(B-2)}$ and $u_2^{(B-2)}$.  Therefore, at
any block $b$ the received signals can be further reduced to:
\begin{align}
y_1^{(b)}=&(f_{11}+{\bf h}_{R1}^{\dagger}{\bf t}_1)u_1'+f_{21}u_2+z_1\label{eq:final2rec3}\\
y_2^{(b)}=&f_{12}u_1+ (f_{22}+{\bf h}_{R2}^{\dagger}{\bf t_2})u_2'+z_2\label{eq:final2rec4}
\end{align}

It is clear that channel does not have the typical form of an interference channel,
\begin{equation}
y_i=h_{ii}x_i+h_{ji}x_j+z_i
\end{equation}
where $i,j \in \{1,2\}$ and $i \neq j$. Hence, we cannot further
reduce the channel to a known form. Each receiver will attempt single
user decoding, i.e. treating interference as noise, and thus can
decode their respective messages $W_1$ and $W_2$ reliably if:
\begin{align}
\label{eq:R1}
R_1\leq \log\bigg(1+&\frac{||f_{11}||^2\psi(P)+||{\bf h}_{R1}^{\dagger}||^2\frac{P_R}{2}}{||f_{21}||^2\chi(P)+1}\nonumber\\
&\qquad+\frac{2 \alpha ||f_{11}||||{\bf h}_{R1}^{\dagger}||\sqrt{\psi(P)\frac{P_R}{2}}}{||f_{21}||^2\chi(P)+1}\bigg)\\
R_2\leq \log\bigg(1+&\frac{||f_{22}||^2\chi(P)+||{\bf h}_{R2}^{\dagger}||^2\frac{P_R}{2}}{||f_{12}||^2\psi(P)+1}\nonumber\\
&\qquad+\frac{2 \alpha ||f_{22}||||{\bf h}_{R2}^{\dagger}||\sqrt{\chi(P)\frac{P_R}{2}}}{||f_{12}||^2\psi(P)+1}\bigg)
\label{eq:R2}
\end{align}
where $\alpha$ equals $0$ when the CSI of the direct links, $f_{11}$
and $f_{22}$, is not available at the relay. The
non-coherent addition of the signals coming from the sources and the
relay which entails a penalty in the achievable rate but does not affect
the DOF. We can set $\alpha=1$ for perfect in-phase addition of the
signals coming from the sources and the relay.\footnote{The
  interested reader can refer to~\cite{Host-Madsen05B} for details on
  the capacity analysis of the full-duplex (a)synchronous relay
  channel with fixed and variable channel gains.}

We proceed to specify power allocation strategies, ranging from very
simple to more sophisticated, and explore the corresponding achievable
degrees of freedom.  Let $\chi(P)=\psi(P)=\frac{P}{2}$ and
$P_R=P$. According to this power allocation, the multi-access part of
the channel according to (\ref{eq:CsMAC}) achieves $DOF=2$.  However,
according to (\ref{eq:R1}) and (\ref{eq:R2}), the signal and
interference have the same power order and hence a $DOF=0$ is
achieved. Therefore, the degrees of freedom of the network in this
case is zero. Clearly this is not a desirable solution.

Now consider an asymmetric power allocation policy characterized by
$\chi(P)=\sqrt{P}$, $\psi(P)=|P-\sqrt{P}|$ and $P_R=P$. In other
words, the cooperative information --also known as the resolution
information-- has a higher power than the information of the current
block of transmission.  It is clear that $DOF=1$ is achieved on the
multi-access side of the channel.  On the other hand, each of
(\ref{eq:R1}) and (\ref{eq:R2}) provides a pre-log factor of
$\frac{1}{2}$ leading to a sum-rate $DOF=1$ for the direct link with
relaying. Therefore, an overall $DOF=1$ is achieved.

\begin{figure}
\centering
\includegraphics{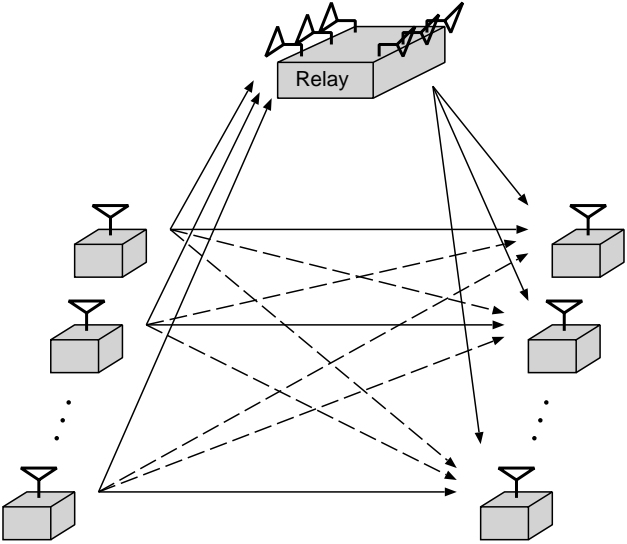}
\caption{Relay for $K$-user interference network.}
\label{fig:network}
\end{figure}
The impact of the above results are clearer when considered in the
context of $K$ users.  The previous coding strategies can easily be
extended to an interference network where $K$ users transmit
simultaneously and a MIMO relay having $M\geq K$ antennas helps all
$K$ nodes in their transmission (see Figure~\ref{fig:network}).

Having a MIMO relay in a network with $K$ source-destination pairs can
have a large impact on the DOF either through a two-hop strategy or
through the coding strategy developed in this section. Specifically
$\frac{K}{2}$ DOF are easily achieved compared to DOF of $1$ with
simple time-sharing strategy in the absence of the relay.

So far, by exploiting the direct links, the DOF are no better than the
simple two-hop strategy. Next, we explore a way to actually capture
the whole $K$ DOF of the channel.

\subsection{Abundant Power at the Relay}

Assume $\chi(P)=\psi(P)=\frac{P}{2}$ at all source nodes while at the
relay we have $P_R=P^2$ (or in general $\mathcal{O} (P^2)$). In this
case, the network will achieve degrees of freedom of $K$, thanks to
the pre-coding strategy employed by the relay, which allows the
relay to avoid causing interference at any node.


\begin{theorem}
The K-users Interference MIMO Relay Channel (IMRC) achieves $K$ (full)
degrees of freedom (per the definition given in (\ref{eq:DOF})), with
the source nodes having each a per block power constraint of $P$ and
the MIMO relay having a power constraint of $\mathcal{O} (P^2)$ and $M
\geq K$ antennas.
\end{theorem}

Note that our definition of the degrees of freedom in~(\ref{eq:DOF})
concentrates on the power of information-bearing nodes, thus allowing
us to study the effect of abundant power at the relay for this
special case.

Although the $K$ degrees of freedom are achieved with a relay that
enjoys power proportional to $P^2$, it is noteworthy that the required
relay power is {\emph independent on $K$}. 

\section{Numerical Results}
\label{sec:simulations}
We corroborate the analysis by the following numerical example of an
interference MIMO relay channel. The following setup is considered:
\begin{itemize}
\item Two-user channel and the relay has two antennas, i.e., $K=M=2$.
\item The noise variance at all nodes $\sigma^2$=1.
\item The magnitude of
channel coefficients are selected as: $f_{11}=2$, $f_{12}=0.75$, $f_{21}=0.75$,
$f_{22}=2$, ${\bf g}_{1R}^\dagger=[2\;\;0.8]$, ${\bf
h}_{2R}^\dagger=[1.2\;\;2]$, ${\bf h}_{R1}^\dagger=[2\;\;1]$ and ${\bf
h}_{R2}^\dagger=[1\;\;0.8]$.
\end{itemize}
An interference channel, without the MIMO relay, can be transformed
into a well known form in the literature known as the standard form
(see {\it e.g.}~\cite{Kramer06}).  The original interference channel
has the following standard form channel gains, under the assumption of
unity noise variance at all nodes,
\begin{align}
a_{12}=\bigg(\frac{f_{12}}{f_{11}}\bigg)^2,\nonumber\\
a_{21}=\bigg(\frac{f_{21}}{f_{22}}\bigg)^2,
\end{align}

The above values of the channel coefficients results in $a_{12}=a_{21}
=0.14$. Thus, without the MIMO relay, the interference is considered
weak/moderate. This is the case where the capacity region of the
interference channel is unknown and where a form of relaying will be
of greater impact on the capacity~\cite{Host-Madsen05}.
\begin{figure*}
\centering
\includegraphics{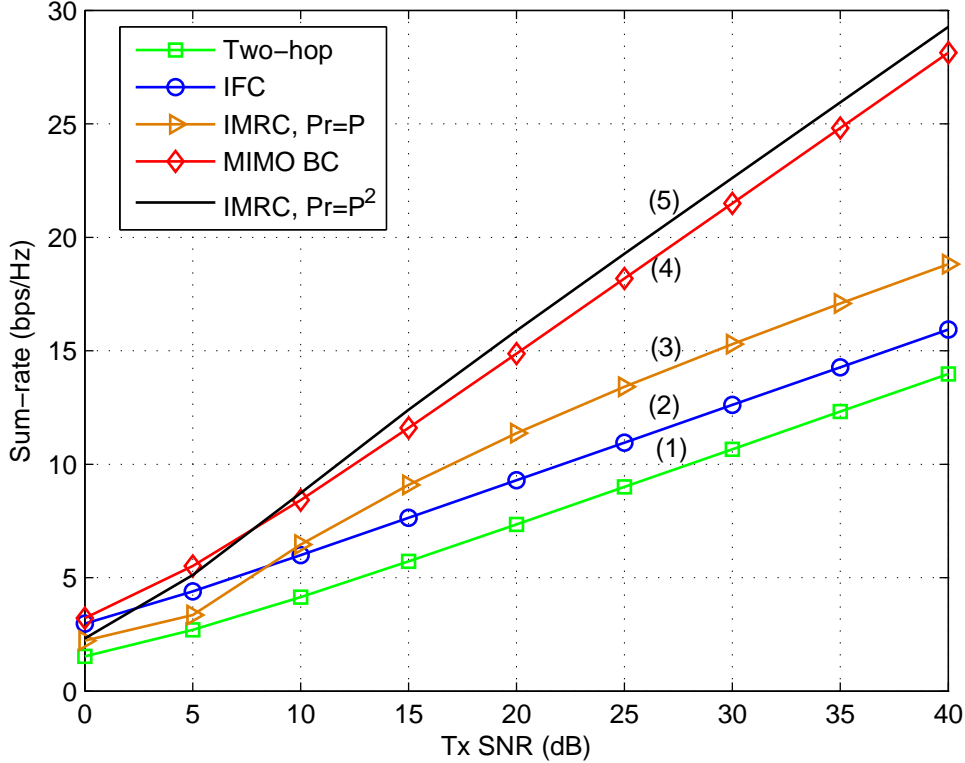}
\caption{Network throughput Vs. transmit SNR $P/\sigma^2$ of each source:
  2-user example.}
\label{fig:sum-rate}
\end{figure*}

Figure~\ref{fig:sum-rate} depicts the sum-rate of five schemes.
Curve~$(1)$ is the sum-rate with a simple two-hop scheme.  Curve~$(2)$
depicts the best known achievable sum-rate for the interference
channel (IFC), with no relay present, using the Han-Kobayashi coding
scheme. This scheme involves rate splitting, joint decoding at the
receivers and moreover it includes a time-sharing random variable that
switches between time-division transmission and simultaneous
transmission by the source nodes. The cardinality of the time-sharing
parameter is set to two and furthermore the power allocation of the
rate-splitting scheme is optimized. This corresponds to curve $4$
of~\cite{Sason04}.  Curve $(3)$ is the computable sum-rate of the
interference MIMO relay channel (IMRC) with the coding strategy
discussed in Section~\ref{sec:coding} and under the asymmetric power
allocation policy characterized by $\chi(P)=\sqrt{P}$,
$\psi(P)=|P-\sqrt{P}|$ and $P_R=P$. Curve~$(4)$ is the case where no
relay is present and the two sources have ideal cooperation leading to
a MIMO BC model. The optimal power allocation and hence the sum-rate
capacity are computed according to {\it Algorithm 2}
of~\cite{Jindal05}. Finally, Curve~$(5)$ is the sum rate of the IMRC,
however, under the assumption of abundant power at the relay,
specifically, $\chi(P)=\psi(P)=\frac{P}{2}$ at the source nodes while
at the relay we have $P_R=P^2$.

We emphasize here that for IMRC, we use independent decoding at the
nodes and we do not fully optimize the power allocation strategies at
the sources and the relay. The focus of this paper is on the DOF and
thus the throughput optimization and analysis at finite SNR is outside
the scope of this paper.

The following conclusions can be drawn from the figure:
\begin{itemize}
\item The three lower curves share the same slope as the corresponding
  schemes have a $DOF=1$. The upper two curves also share another
  slope verifying that the IMRC with $P_R=P^2$ achieves the full DOF
  of the channel.
\item It is interesting to see that the fully optimized IFC as
  explained above can achieve higher throughput than the simple
  two-hop scheme; the fully optimized IFC in this figure has a total
  power equal to the total power of all the nodes in the IMRC,
  including the relay. The simple two-hop scheme does not involve
  optimal power allocation at the source nor joint decoding at the
  destinations.
\item While the coding strategy devised in this paper to exploit the
  direct links of the IMRC of Curve~$(3)$ shares the same DOF of IFC,
  but as it can be seen in the figure, this strategy leads to
  noticeable gains in the sum-rate for medium and high SNR values over
  the IFC.
\item The IMRC scheme with abundant power at the relay, Curve~$(5)$,
  presents a substitute strategy for the cooperative MIMO,
  Curve~$(4)$. Cooperative MIMO is a scheme of interest in the
  wireless industry to improve the throughput for the uplink of
  cellular networks.
\end{itemize}

\section{Discussion}
\label{sec:conc}

In this work, we characterize the high-SNR sum-rate behavior of an
interference channel with a MIMO relay. First, we establish that the
degrees of freedom of the interference MIMO relay channel to be only
$\frac{K}{2}$. Therefore, in this case exploiting the direct links
does not provide significant throughput enhancement at high SNR.
Then, we consider a more sophisticated coding scheme that exploits the
direct links and the possibility of abundant power at the relay. We
show that if the transmit/receive nodes have power proportional to
$P$, and the relay has power proportional to $P^2$, all $K$ degrees of
freedom of the channel become available. This result is achieved under
modest channel knowledge assumption at the network nodes and with the
assumption of single user decoding at the destinations.

While we consider the case of full-duplex relay, one can devise
similar signaling strategies for the half-duplex case. However, the
block-Markov coding is not required. A brief description of a possible
coding scheme is given as follows. The sources transmit all the
time. However, they divide each block of their transmission into two
halves.  Each source node transmits the same message in the two halves
using i.i.d. Gaussian code books. During the second half, the relay
transmits and manages the interference as discussed above in the
full-duplex case. At the destinations, the received signals at the
first and second halves form two Gaussian parallel channels, the first
sees interference while the other is interference-free. It can be
easily shown that the maximum degrees-of-freedom of this scheme is
$\frac{K}{2}$. As we know from {\it Theorem \ref{theorem:DOF}}, a
two-hop strategy suffices to achieve a $\frac{K}{2}$
degrees-of-freedom. However, exploiting the direct links provides an
increase in the throughput compared to two-hop communications for all
signal-to-noise-ratios.

Several directions naturally arise for future work. Our analysis
concentrates on the high SNR behavior of the network throughput, thus
many parameters of the IMRC can be further optimized for non-asymptotic
SNR values. More complex coding/decoding techniques can also be employed, for
example, a modified Han-Kobayashi scheme (in the presence of the MIMO relay)
that combines rate-splitting, time-sharing (TDM), relaying and joint decoding
at the receivers.

Finally, it is worth mentioning that since the time of submission of
this paper for publication there has been a surge of research
activity in the area of interference networks. In particular, the
time/frequencey non-selective interference channel has been studied
lately in more depth and a better understanding of its performance
limits has been developed. The interested reader can refer
to~\cite{Etkin09,Khandani09} and the references therein.

\appendix
\label{app:BFvectors}

For simplicity we consider the two user case. Denote by ${\bf
  t}_1=[t_{11} \;\; t_{12}]$ and ${\bf t}_2=[t_{21} \;\; t_{22}]$ the
beamforming vectors at the relay.

The following conditions govern the selection of the beamforming
vector ${\bf t}_1$ ,
\begin{enumerate}
\item $||{\bf t}_1||^2=\frac{P_R}{2\psi(P)}$
\item ${\bf h}_{R2}^{t} {\bf t}_1=-f_{12}$
\item $\Phi_1=\angle f_{11}=\angle {\bf h}_{R1}^{t} {\bf t}_1$
\end{enumerate}
where $\Phi_1$ is the phase of the (direct) channel between source
node $1$ and the intended destination node. Similarly, One can write
the conditions for selecting ${\bf t}_2$.

The first condition is related to power scaling at the relay. The
second condition is the one responsible to reduce the interference
seen by the destination nodes. The third condition is optional, it is
responsible for coherent combination of the desired signal component
at the intended destination. It requires though global channel
knowledge at the relay.

A closed form solution for the three simultaneous conditions is not
feasible. Instead we assume the lack of the third condition which does
not affect the DOF achieved by the coding scheme for IMRC explained in
this paper and simplifies the channel knowledge requirements.

Solving for ${\bf t}_2$ and ${\bf t}_2$, one gets for $m \neq n$, and
$m, n \in \{1, 2\}$,

\begin{equation}
t_{m1}=\frac{\pm h_{Rn,2}\sqrt{-f_{m2} \pm ||{\bf h}_{Rn}||^2\frac{P_R}{2\psi(P)}}-f_{m2}{\bf h}_{Rm,1}}{||{\bf h}_{Rn}||^2}
\end{equation}
and,
\begin{equation}
t_{m2}=-\frac{f_{m2}}{h_{Rn,2}}-\frac{h_{Rn,1}}{h_{Rn,2}}t_{m1}
\end{equation}

\bibliographystyle{IEEEtran}
\bibliography{IEEEabrv,ramy}

\end{document}